\documentclass[12pt, twoside]{article} 
\usepackage{amsmath} 
\usepackage{amscd} 
\usepackage{amsfonts} 
\usepackage{amsthm} 
\usepackage{latexsym} 
\theoremstyle{definition} 
\newtheorem{defn}{Definition}[section] 
\theoremstyle{plain} 
\newtheorem{thm}{Theorem}[section] 
\theoremstyle{remark} 
\newtheorem{rem}{Remark}[section] 
\newtheorem{exmp}{Example}[section] 
\newcommand{\field}[1]{\mathbb{#1}} 
\newcommand{\F}{\field{F}} 
\newcommand{\R}{\field{R}}

\newcommand{\bo}[1]{\boldsymbol{#1}} 
\newcommand{\mf}[1]{\mathsf{#1}} 
\newcommand{\mc}[1]{\mathcal{#1}} 
\begin{document} 
\title{A linear solution to the effective four-dimensionality problem} 
\author{Vladimir Trifonov\\AMS\\trifonov@member.ams.org \\
Updated version of \\ \emph{A Linear Solution Of The Four-Dimensionality Problem} \\ 
Europhysics Letters $\bo{32 (8)}$, 1995, pp. 661-626} 
\date{}\maketitle 
\begin{abstract} 
In this note we formalize certain aspects of observation process in an attempt to link the logic of the observer with properties of the observables structures. It is shown that an observer with Boolean logic perceives her environment as a four-dimensional  Lorentzian manifold. 
\end{abstract} 

\section*{Introduction.} Topos theory (\cite{AGV72}, \cite{La64}, \cite{Mi72}, 
\cite{Co73}, \cite{Jo77}, \cite{ Go84}) offers an independent (of the set theory) 
approach to the foundations of mathematics. Topoi are categories with set-like objects, 
function-like arrows and Boolean-like logic algebras. Handling these generalized sets and 
functions in a topos may differ from that in classical mathematics (i.e. the topos 
$\bo{Set}$ of sets): there are non-classical versions of mathematics, each with its 
non-Boolean version of logic. One possible view on topoi is this: abstract worlds, 
universes for mathematical discourse, inhabitants \emph{(observers)} of 
which may use non-Boolean logics in their reasoning. From this viewpoint the main 
business of classical physics is to construct models of the universe with a given 
\emph{bivalent Boolean} model of the observer, and choose the most adequate one. In a 
sense, our task is inverse: with a given model of the universe, to construct models of the 
observer, and find out how the observer's perception of the universe changes if his logic 
is changed. Thus, not the universe itself, but rather its \emph{differential} is what interests us here. 
\begin{rem} For structural clarity we use $\Box$ at the end of a \emph{proof}, and each \emph{example} ends with $\triangle$. To conclude a \emph{remark} we use the diamond sign. \hfill $\Diamond$ \end{rem} 

\section{Actions.} We start by describing the observer's interactions (\emph{actions} and 
\emph{observations}) with the environment. The major intuition-based attribute of actions (elementary influences of the observer upon the environment) is that they can be associatively composed (i.e. performed in sequence), the compositions also being actions, and there is an \emph{identity} action (changing nothing). The set of the observer's actions (or \emph{effectors}), together with an associative composition, is his \emph{motor domain}. 
\begin{exmp} In quantum theory the observer's actions are represented by operators on a 
linear space and constitute, together with an associative composition, a semi-group with an 
identity (monoid). \hfill $\triangle$ \end{exmp} 

\section{Observations.} The major intuition-based property of observations 
(mental and visual pictures of fragments of reality and appearance) is their ability to be 
superposed, with some real (later we shall generalize the situation for an arbitrary field 
$\F$) weight factors, assigned by the observer to each item. Intuitively, they measure 
the participatory degree of observations in a particular observational situation. Formally, 
there are two algebraic operations on the set of observations: addition and multiplication 
(by reals). The set of observations (or \emph{reflexors}), together with the two operations, is the observer's \emph{sensory domain}. 
\begin{exmp} Spacetime $S$ of special relativity can be interpreted as the set of 
observations (mental and visual images of events) of an observer: he considers nearby 
events as superpositions of some observations taken with some real weight factors 
(decomposition of an event in a basis). Since $S$ is a real linear space, there are, indeed, 
two operations on it - addition and multiplication by reals. \hfill $\triangle$ \end{exmp} 
\section{Combining actions with observations.} \emph{No elementary phenomenon is a phenomenon until it is an observed} \emph{phenomenon} (observations are actions). This simply means that obtaining constructive information about reality changes its appearance. Observations are combined with actions into new entities, called by constructivists \emph{states of knowledge}: any rational observer performs an action in accordance to, and interprets an observation on the basis of, his particular state of knowledge. It is said that rational knowledge consists of two fundamental (\emph{sensory} and \emph{motor}) components. 
\section{Paradigm.} We shall call the set of a observer's states of knowledge his 
\emph{paradigm}. Observations, then, induce superposition of states of knowledge  (\emph{extensive} acquisition or \emph{accumulation} of knowledge), 
and actions induce associative composition (\emph{intensive} acquisition or 
\emph{elevation} of knowledge). Taking into account weight factors, we get another 
operation, the \emph{protensive} acquisition of knowledge. Thus we have three operations 
on the observer's paradigm, which endows it with an algebraic structure. The linear case of 
this structure is, of course, a real linear associative algebra $A$ with an identity. The 
sensory domain $S_A$ then is the additive linear space of the algebra, and the motor 
domain $M_A$ is, one would say, its multiplicative monoid $M$. However, it is not terribly easy to interpret $\bo{0}$ (the zero of the algebra) as an action. The identity $\bo{1}$ of the algebra is the identity action, but what is $\bo{0}$? We would rather take $M \setminus \bo{0}$ as the motor domain, but in the former a composition of two actions is 
not always an action (i.e. $M \setminus \bo{0}$ is not always a monoid), which violates 
the intuitive notion of action and, moreover, will not let us define the logic of the observer. To make a compromise, we assume that $M_A = M \setminus \bo{0}$ if the latter is a 
monoid, otherwise $M_A = M$. In other words, the motor domain is the monoid generated by the set of non-zero elements of $M$. 
\section{Comprehension signature.} The operations defining the paradigm can be 
described by its (algebraic) \emph{signature}, i.~e., the set of operation 
symbols $\Sigma := \{+, \cdot, \bo{0}, \bo{1}\} \cup \R$, each with its \emph{arity}, a 
natural number. For example the arity of the symbols + and $\cdot$ is 2, while $\bo{0}$ 
and $\bo{1}$ both have arity $0$ (they are \emph{constants}). Each real number $r \in 
\R$ is an operation symbol of arity $1$. The algebraic signature together with a set of 
equations specifying behavior of linear algebras over a field $\F$ is referred to as the 
\emph{comprehension} \emph{signature} of the observer. 
\section{Time.} We employ the constructivistic concept of time: a fundamental attribute 
of thought process, the basis to distinguish one entity from another. No statement on time 
being a physical property of the universe is made. Constructivists describe time as a partial order on the set of states of knowledge. So do we, slicing the paradigm with a one-form $\tilde{\tau}$ (\emph{time gradient} or \emph{ether form}) on its sensory domain, which partially orders states of knowledge by the standard ordering of the set $\R$ of reals. 
\section{Metric.} An observer's natural ability to estimate angles and distances between 
observations is represented by a metric on his sensory domain $S_A$. We do not force 
metrics into the scheme because a natural metric is defined automatically, once the ether  
form of the observer is known, as follows. Each real algebra $A$ is completely defined 
by the structure constant tensor $\bo{\mf{A}}(\tilde{\omega}; a, b)$ on its additive linear 
space $S_A$. Tensor $\bo{\mf{A}}$ is a multilinear function of two vector arguments $a, 
b$ and a one-form argument $\tilde{\omega}$. Choosing a particular one-form $\tilde{\tau}$ (i. e. ether form) on $S_A$ makes the tensor $\bo{\mf{A}}(\tilde{\tau}; a, b)$ depend only on the vector arguments. Thus, if $\bo{\mf{A}}(\tilde{\tau}; a, b)$ is symmetric in $a$ and $b$, it is a (proper- or pseudo-)Euclidean metric on $S_A$. 
\section{The environment.} We assume that the \emph{universe} consists of \emph{interacting systems}. Each system is represented by its \emph{states}. Given all states of a system, it is defined completely. Some different systems may have the same states (\emph{common} states). A system $X$ is a \emph{subsystem} of a system $Y$ if all states of $X$ are common to $X$ and $Y$. If two systems are subsystems of each other, it is natural to consider them equal. Given two systems $X$ and $Y$, we can consider a system $Z$ (the \emph{union} $X$ and $Y$ whose states are all states of $X$ and all states of $Y$. For 
two systems $X$ and $Y$ with common states there is a system $Z$ (the \emph{intersection} of $X$ and $Y$ whose states are their common states. A system $X$ that can have only states that $Y$ cannot, is the \emph{complement} of $Y$. The behavior of the ontological pair $(system, state)$ resembles that of $(set, element)$ in na\"{\i}ve set theory, although conceptually they are very different. Two systems \emph{interact} if states of one system depend on states of the other one, which is described as a \emph{function} in set-theoretic terms. Thus, with systems as sets and interactions as functions, the category $\bo{Set}$ serves as a first-order model of the environment axioms. 
\section{The proper universe.} The observer's actions change states of a system i.e. any 
action $a$ induces a map $X \to X$, and we have the influence of the observer with the 
motor domain $M$ on a system $X$ as a realization of the monoid $M$ in the set $X$, 
i.e. a map $\lambda$, assigning to each $a \in M$ a function $f_a: X \to X$ such that: $f_a 
\circ f_b = f_{ab}$, and  $f_{e}$ is the identity map i.e. $f_{e}(x) = x, \forall x\in X, 
e$ is the identity of $M$. A pair $(X, \lambda)$, where $\lambda$ is a realization of 
a monoid $M$ in a set $X$, is called an $M$-system. The collection $M\bullet\bo{Set}$ 
of all $M$-systems describes all possible influence of a observer with the motor domain 
$M$ on the universe. $M\bullet\bo{Set}$ is a topos in which arrows $(X, 
\lambda) \to (Y, \mu)$ are realization preserving maps $f: X \to Y$, i.e. such, that the 
following diagram commutes for each $a \in M$: 
\begin{displaymath} \begin{CD} X @>f>> Y\\ 
@V\lambda_aVV @VV\mu_{a}V\\ X @>f>> Y \end{CD}. \end{displaymath} 
The principle of \emph{active} \emph{comprehension} (the logic of an observer is developed in his interaction with the environment) defines the \emph{proper} \emph{universe} as the topos $M\bullet\bo{Set}$ and assigns to the observer its logic and 
mathematics. 
\section{Technical setup.} The intuition-based concepts we used above are logically prior to physics. Each is elementary in the sense that it is just several steps from the set and category axioms. To compare, the notion of smooth manifold (a starting point for the working 
physicist) is far more complicated. Let us describe the above outline a bit more precisely. 
\begin{defn} Let $\F$ be a partially ordered field, called a \emph{protensity}. For a protensity $\F$, the (\emph{Boolean}) $\F$-\emph{observer} is the 
category $A[\F]$ of linear algebras over $\F$, with objects called \emph{paradigms} and 
arrows called \emph{shifts}. For a paradigm $A$ \emph{states} \emph{of} 
\emph{knowledge} are elements of the algebra $A$. The additive linear space $S_A$ of the algebra $A$ is the \emph{sensory domain}, with elements called \emph{reflexors}. The dual space $S^{\ast}_A$ is the \emph{temporal domain} with elements called \emph{ether forms}. \end{defn} 
\begin{rem} Intuitively, we are to think of $\F$ as the \emph{syntax} of the observer's    \emph{language} of \emph{thought}. \hfill $\Diamond$ \end{rem} 
\begin{defn} For a paradigm $A$, a \emph{principal metric} is the structure 
constant tensor $\bo{\mf{A}}$ together with an ether form $\tilde{\tau}$, provided $\bo{\mf{A}}(\tilde{\tau}; a, b)$ is symmetric in vector arguments $a$ and $b$. \end{defn} 
\begin{rem} A paradigm may have several metrics or it may have none. \hfill $\Diamond$ 
\end{rem} 
\begin{defn} The \emph{motor domain} is the multiplicative subgroupoid $M_A$ of the 
algebra $A$, generated by the set of all non-zero elements of $A$. The elements of $M_A$ are referred to as \emph{effectors}. A paradigm $A$ is \emph{(ir)rational} if $M_A$ is (not) a monoid. If $A$ is a rational paradigm and the topos of $M_A$ realizations, $M_A\bullet\bo{Set}$, is Boolean, the paradigm $A$ is \emph{consistent}. The topos $M_A\bullet\bo{Set}$ is the \emph{proper} \emph{universe} (or the \emph{monocosm}) of the paradigm $A$. \end{defn} 
\begin{rem} The environment ($\bo{Set}$) is a topos of realizations of a single-element monoid, therefore it is the proper universe of an \emph{absolutely objective} paradigm whose 
motor domain contains the identity action only. Informally, any absolutely objective 
observer is absolutely inert. \hfill $\Diamond$ \end{rem} 
\begin{defn} A consistent paradigm of maximal finite dimensionality, if it exists, is a 
\emph{home} paradigm. \end{defn} 
\section{$\R$-observer.} We now apply this scheme to the Boolean $\R$-observer. The conclusion we shall obtain is that $S_A$ is Minkowski space. 
\begin{thm} For each home paradigm of the Boolean $\R$-observer, its sensory domain is four-dimensional, and each principal metric is Lorentzian. \end{thm} 
\begin{proof} If, for a paradigm $H$, the logic of the topos $M_H\bullet\bo{Set}$ is Boolean then $M_H$ is a group (\cite{Go84}, p. 121). Therefore $H$ is associative, with an identity and without zero divisors. Any such $H$ is isomorphic to the quaternion algebra (see \cite{Ku73}). Hence $\bo{0} \not\in M_H$ ($\bo{0}$ has no inverse) i.~e., the motor domain $M_H$ is isomorphic to the group of nonzero quaternions $\mc{H}$. Thus a home paradigm 
exists, and any such paradigm is four-dimensional. For a basis $e_{\beta}$ in $S_H$ let 
$t_{\beta}$ be the components of a one-form $\tilde{\tau}$ in the dual basis $e^{\beta}$ (the indices run from 0 to 3). Then components $G_{\alpha \beta}$ of the principal metric $\bo{G} = \bo{\mf{H}}(\tilde{\tau}; a, b)$ are 
\begin{equation} G_{\alpha \beta} = \bo{\mf{H}}(\tilde{\tau}_{\gamma}e^{\gamma}; 
e_{\alpha},e_{\beta}) = \tilde{\tau}_{\gamma}\bo{\mf{H}}(e^{\gamma}; e_{\alpha}, e_{\beta}) = 
\tilde{\tau}_{\gamma}\mf{H}^{\gamma}_{\alpha \beta}, \end{equation} 
where $\mf{H}^{\gamma}_{\alpha\beta}$ are the components of $\bo{\mf{H}}$. They are easily found in the canionical basis of unit quaternions $\bo{1}$, $\bo{i}$, $\bo{j}$, $\bo{k}$: 
\begin{displaymath} G_{\alpha \beta} = \begin{pmatrix} 
\tilde{\tau}_0& \tilde{\tau}_1& \tilde{\tau}_2& \tilde{\tau}_3\\ 
\tilde{\tau}_1&-\tilde{\tau}_0& \tilde{\tau}_3&-\tilde{\tau}_2\\ 
\tilde{\tau}_2&-\tilde{\tau}_3&-\tilde{\tau}_0& \tilde{\tau}_1\\ 
\tilde{\tau}_3& \tilde{\tau}_2&-\tilde{\tau}_1&-\tilde{\tau}_0 \end{pmatrix}. \end{displaymath} 
Since it must be symmetric, we demand nontrivial symmetry $\tilde{\tau}_1=-\tilde{\tau}_1$, $\tilde{\tau}_2=-\tilde{\tau}_2$,  $\tilde{\tau}_3=-\tilde{\tau}_3$, which yields $\tilde{\tau}_1=\tilde{\tau}_2=\tilde{\tau}_3=0$. Thus for each $\tau \in \R \setminus \{0\}, H$ has a principal metric of signature 2, generated by an ether form $\tilde{\tau}$ with the components $(\tau, 0, 0, 0)$ in the canonical basis, which concludes the proof. \end{proof} 
\begin{rem} Once the ether form $\tilde{\tau}$ generates a metric, the latter in turn generates time in its standard sense. If we ignore the motor structure of the paradigm, four-dimensionality and Lorentz metric become axioms, which is the case in both quantum mechanics and relativity. \hfill $\Diamond$ \end{rem} 
\section{Spacetime.} Since the environment $\bo{Set}$ is a Boolean topos, we can separate a 
Boolean part - the most objective, in a sense, in any rational paradigm $A$. A natural 
candidate is the group $\mc{A}$ of invertible elements of $A$. For any 
finite-dimensional rational paradigm $A$ of $\R$-observer, $\mc{A}$ is a Lie group. 
The sensory domain can be identified with the tangent space at the identity $e$ of $\mc{A}$, $S_A \cong T_e\mc{A}$. If $A$ has a principal metric $\bo{G}$, then it can be naturally extended over $\mc{A}$ in a number of ways, endowing it with a (pseudo- or  proper-)Riemannian structure (see \cite{Tr03} for the explicit computations), so it can be considered as the (observable) spacetime of the paradigm $A$.

\end{document}